\documentclass[11pt]{article}
\usepackage{color,amsthm,amsmath,amsfonts,xspace}
\usepackage{MnSymbol}
\usepackage{fullpage}
\usepackage{hyperref}
\usepackage{authblk}

\newtheorem{Theorem}{Theorem}
\newtheorem{Theorem*}{Theorem}

\newtheorem{Claim*}[Theorem]{Claim}
\newtheorem{Corollary}[Theorem]{Corollary}

\newtheorem{CounterExample*}{$\overline{\hbox{\bf Example}}$}

\newtheorem{Example*}[Theorem]{Example}

\newtheorem{Intuition*}[Theorem]{Intuition}
\newtheorem{Joke*}[Theorem]{Joke}
\newtheorem{Lemma}[Theorem]{Lemma}
\newtheorem{Lemma*}[Theorem]{Lemma}
\newtheorem{Open problem}[Theorem]{Open problem}

\newtheorem{Question*}[Theorem]{Question}
\newtheorem{Remark}[Theorem]{Remark}

\def \bSubexa    {\begin{subexa}}

\newcommand{\ignore}[1]{}

\newcommand{\EE}{\mathbb{E}}

\newcommand{\NN}{\mathbb{N}}

\newcommand{\RR}{\mathbb{R}}
\newcommand{\ZZ}{\mathbb{Z}}

\newcommand{\integers}{\ZZ}
\newcommand{\naturals}{\NN}

\newcommand{\reals}{\RR}

\newcommand{\integersp}{\integers^+}

\def \cA     {{\cal A}}

\def \cD     {{\cal D}}
\def \cE     {{\cal E}}

\def \cM     {{\cal M}}

\def \cP     {{\cal P}}

\def \cX     {{\cal X}}
\def \cY     {{\cal Y}}

\newcommand{\eg}{\textit{e.g.,}\xspace}
\newcommand{\ie}{\textit{i.e.,}\xspace}  
\newcommand{\iid}{\textit{i.i.d.}} 

\definecolor{light}{gray}{.75}

\def \eqed    {\eqno{\qed}}

\def \upto  {{,}\ldots{,}}

\def \sets#1{{\{#1\}}}
\def \Sets#1{{\left\{#1\right\}}}

\def \floor#1{{\lfloor{#1}\rfloor}}

\def \paren#1{{({#1})}}
\def \Paren#1{{\left({#1}\right)}}

\def \binomial#1#2{{{#1}\choose{#2}}}

\newcommand{\ed}{\stackrel{\mathrm{def}}{=}}

\def\ignore#1{}

\newcommand{\bi}{\begin{itemize}}
\newcommand{\ei}{\end{itemize}}

\def\orpro{\mathop{\mathchoice
   {\vee\kern-.49em\raise.7ex\hbox{$\cdot$}\kern.4em}
   {\vee\kern-.45em\raise.63ex\hbox{$\cdot$}\kern.2em}
   {\vee\kern-.4em\raise.3ex\hbox{$\cdot$}\kern.1em}
   {\vee\kern-.35em\raise2.2ex\hbox{$\cdot$}\kern.1em}}\limits}

\def\andpro{\mathop{\mathchoice
 {\wedge\kern-.46em\lower.69ex\hbox{$\cdot$}\kern.3em}
 {\wedge\kern-.46em\lower.58ex\hbox{$\cdot$}\kern.25em}
 {\wedge\kern-.38em\lower.5ex\hbox{$\cdot$}\kern.1em}
 {\wedge\kern-.3em\lower.5ex\hbox{$\cdot$}\kern.1em}}\limits}

\def\simge{\mathrel{%
   \rlap{\raise 0.511ex \hbox{$>$}}{\lower 0.511ex \hbox{$\sim$}}}}

\def\simle{\mathrel{
   \rlap{\raise 0.511ex \hbox{$<$}}{\lower 0.511ex \hbox{$\sim$}}}}

\DeclareMathOperator*{\argmax}{arg\,max}

\DeclareMathOperator*{\union}{\cup}

\newcommand{\cXn}{\cX^n}

\newcommand{\bE}{\EE}
\newcommand{\bEP}{\operatorname*{\bE}_P\;}

\newcommand{\as}{k}
\newcommand{\lng}{n}

\newcommand{\nlower}{\lng_1}

\newcommand{\param}{\lambda}

\newcommand{\distP}{P}
\newcommand{\distPn}{\distP^n}
\newcommand{\distPnp}{\distP^{\np}}
\newcommand{\distPPsnn}{\distP^{\Psnn}}

\newcommand{\distQ}{Q}

\newcommand{\dclass}{\cP}
\newcommand{\dclassn}{\dclass^{\lng}}
\newcommand{\dclasspsnn}{\dclass^{{\Psns\lng}}}

\newcommand{\symb}{x}
\newcommand{\Symb}{X}
\newcommand{\symby}{y}

\newcommand{\wrpn}{\wred(\dclassn)}

\newcommand{\varx}{X}

\newcommand{\xn}{x^n}
\newcommand{\xnp}{x^{\np}}
\newcommand{\Xn}{X^n}

\newcommand{\distqn}{\distQ_n}

\newcommand{\barF}{\overline{F}_u}

\newcommand{\red}{R}
\newcommand{\wred}{\hat{R}}
\newcommand{\ered}{\overline{R}}

\newcommand{\cDk}{\cD_k}
\newcommand{\cDkn}{\cDk^n}
\newcommand{\cDkpsnn}{\cDk^{\Psnn}}
\newcommand{\cDkpsnnp}{\cDk^{\Psnnp}}

\newcommand{\cDinf}{\cD_\infty}

\newcommand{\Envelope}{\cE}
\newcommand{\Envcls}{\Envelope_f}
\newcommand{\Envclsi}{\Envelope_{f_i}}
\newcommand{\Envclsn}{\Envcls^n}
\newcommand{\Envclspsnn}{\Envcls^{\Psnn}} 

\newcommand{\Expcalp}{\Envelope_{c\cdot e^{\alpha i}}}
\newcommand{\Expcalpn}{\Expcalp^n}
\newcommand{\Expcalppsnn}{\Expcalp^{\Psnn}}

\newcommand{\Pwrcalp}{\Envelope_{c\cdot i^{-\alpha}}}
\newcommand{\Pwrcalpn}{\Pwrcalp^n}

\newcommand{\mon}{\cM}

\newcommand{\monkn}{\mon_k^n}

\newcommand{\shtarkov}{S}

\newcommand{\mlp}{\hat{\distP}}
\newcommand{\mlpx}{\mlp(\symb)}
\newcommand{\mlpxn}{{\mlp}^n(\xn)}
\newcommand{\mlpn}{\hat{\distPn}}

\newcommand{\const}{c}

\newcommand{\upper}{{\Lambda}}

\newcommand{\slower}{l}
\newcommand{\trans}{b}

\newcommand{\type}{\tau}

\newcommand{\tcr}[1]{\textcolor{red}{#1}}

\newcommand{\psnfrm}[2]{e^{-{#1}}\cdot{\frac{{#1}^{#2}}{#2!}}} 

\newcommand{\Poisson}{\rm poi}

\newcommand{\Psns}[1]{{\Poisson}({#1})} 
\newcommand{\Psnlmb}{\Psns\lambda}
\newcommand{\Psni}{\Psns i}
\newcommand{\Psnn}{\Psns n}
\newcommand{\Psnnp}{\Psns {n'}}
\newcommand{\psnarg}{i}

\newcommand{\np}{n'}
\newcommand{\Np}{N}

\newcommand{\Psnss}[2]{{\Poisson}({#1},{#2})} 
\newcommand{\Psnlmbarg}{\Psnss\lambda\psnarg}
\newcommand{\Psnlmbi}{\Psnss\lambda i}
\newcommand{\Psnnnp}{\Psnss n{n'}}

\newcommand{\cPPsns}[1]{{\cal{POI}}_{#1}}
\newcommand{\cPPsnupr}{\cPPsns\upper}

\newcommand{\mlt}{m}
\newcommand{\Mlt}{M}
\newcommand{\MltPsnn}{\Mlt^{\Psnn}}

\newcommand{\vecx}{\overline{x}}

\renewcommand{\iid}{\textit{i.i.d.}\xspace}
\title{
Universal Compression of Envelope Classes:\\
Tight Characterization via Poisson Sampling\thanks{A part of the paper
will appear in IEEE International Symposium on Information Theory 2014}}
\author[1]{Jayadev Acharya\thanks{ jayadev@csail.mit.edu. Part of the
    work done while author was
a student at UCSD.}}
\author[2]{Ashkan Jafarpour\thanks{ashkan@ucsd.edu}}
\author[2]{Alon Orlitksy\thanks{alon@ucsd.edu}}
\author[2]{Ananda Theertha Suresh\thanks{asuresh@ucsd.edu}}
\affil[1]{Massachusetts Institute of Technology}
\affil[2]{University of California, San Diego}
\begin{document}
\maketitle
\begin{abstract}
The Poisson-sampling technique eliminates dependencies among
symbol appearances in a random sequence. It has been used to
simplify the analysis and strengthen the performance guarantees
of randomized algorithms.
Applying this method to universal compression,
we relate the redundancies of fixed-length and Poisson-sampled sequences,
use the relation to derive a simple \emph{single-letter formula}
that approximates the redundancy of \emph{any} envelope class to within
an additive logarithmic term. As a first application, we consider \iid
distributions over a small alphabet as a step-envelope class, and provide
a short proof that determines the redundancy of discrete distributions
over a small alphabet up to the first order terms. We then show the
strength of our method by applying the formula to tighten the existing
bounds on the redundancy of exponential and power-law classes,
in particular answering a question posed by Boucheron, Garivier and Gassiat~\cite{BGG09}.
\end{abstract}

\section{Introduction}
Compression concerns efficient representation of random phenomena.
Let a random variable $X$ be generated according to a distribution
$\distP$  over a discrete set $\cX$.
The best compression of $X$ is achieved when $\distP$ is known in
advance.
Roughly speaking, every $\symb\in\cX$ is then represented by 
$\log(1/\distP(\symb))$ bits, where throughout the paper $\log$
represents logarithm to the base 2 and $\ln$ represents the natural
logarithm. Hence the expected number of
bits used is the distribution's \emph{entropy}
\[
H(\distP)
\ed
\sum_{\symb\in\cX}\distP(\symb)\log\frac1{\distP(\symb)}
=
\bEP\log\frac1{P(X)}.
\] 

Universal compression addresses compression in the common setting
where the underlying distribution $\distP$ is not known, but can be
assumed to belong to a known class $\cP$ of distributions over $\cX$,
for example the class of \iid or Markov distributions.
Any compression of $X$ can be viewed as representing $\symb\in\cX$ using
$\log (1/\distQ(\symb))$ bits for some distribution $\distQ$ over
$\cX$~\cite{CT91}, and in the rest of the paper we will use the two interchangeably.
The expected number of bits that $\distQ$ uses to represent $X\sim P$
is the cross entropy
\[
\sum_{\symb\in\cX}\distP(\symb)\log\frac1{Q(\symb)}
=
\bEP \log\frac1{Q(X)}.
\]
The expected additional number of bits, beyond the entropy, that $\distQ$
uses to represent $\Symb$ is 
\[
\bEP\log\frac1{\distQ(\Symb)}-\bEP\log\frac1{\distP(\Symb)}
=
\bEP \log\frac{\distP(\Symb)}{\distQ(\Symb)}.
\] 
Though we will not use this fact, this difference is the KL divergence 
$D(\distP||\distQ)$ between $\distP$ and $\distQ$.


The \emph{expected redundancy} of a distribution collection $\cP$ is
the lowest increase in the expected number of bits achieved by any compression scheme $\distQ$,
\[
\ered(\dclass)
\ed
\inf_{\distQ}\;\sup_{\distP\in\dclass}\;
\bEP \log \frac{\distP(\Symb)}{\distQ(\Symb)}.
\]

A more stringent redundancy measure, which is also the focus of this
paper, is the \emph{worst-case} redundancy (a.k.a. \emph{minimax regret})
that represents the largest possible increase 
between the number of bits that $\distQ$ and the best compression
scheme use to represent any $\symb\in\cX$,
\[
\wred(\dclass)
\ed
\inf_{\distQ}\sup_{\distP\in\dclass}\sup_{\symb\in\cX}
\log \frac{\distP(\symb)}{\distQ(\symb)}
\]

Worst-case redundancy clearly exceeds expected redundancy for 
any distribution collection $\cP$,
\[
\wred(\dclass) 
\ge
\ered(\dclass).
\]
However, for many popular classes of distributions, they are almost the same.

For example, consider the collection of discrete distributions over 
$[k]\ed\sets{1\upto k}$,
\[
\cDk
\ed
\Sets{(p_1\upto p_k): p_i\ge 0, \sum p_i=1}.
\]
It is easy to see that the uniform $Q=(1/k\upto 1/k)$ achieves
redundancy $\log k$, and that any other distribution will have
a higher redundancy, hence
\[
\wred(\cDk) = \ered(\cDk)
=
\log k.
\]

The most well studied distributions are 
\emph{independent identical} distributions, or \emph{iid},
where a distribution $\distP$ over $\cX$ is sampled $n$ times
independently, and the probability of observing $\xn\in\cX^n$ is 
\[
\distPn(\xn)
\ed
\prod_{i=1}^n\distP(x_i).
\] 
Similarly, for a class $\dclass$ define the \iid class
\[
\dclassn
\ed
\{\distPn:\distP\in\dclass\}
\]
to consist of the $n$ independent repetitions of any single
distribution $\distP\in\dclass$.

The most well investigated class of distributions is $\cDkn$,
the collection of all length-$n$ \iid distributions over $[k]$.

It is now well established~\cite{Kie78, Dav73,DMPW81,WST95,Cov91,Ris96,Szp98,XB00,OS04,SW10} 
that for $k= o(n)$
\[
\ered(\cDkn) +f_1(k)
=
\wred(\cDkn) +f_2(k)
=
\frac{k-1}2\log \frac {\lng}{\as},
\]
and for $\lng = o(\as)$
\[
\ered(\cDkn)+g_1(n)=\wred(\cDkn)+g_2(n) = n\log \frac {\as}{n},
\]
where $f_1,f_2$ are independent of $n$ and $g_1,g_2$ are independent of $k$. 

The problem was traditionally studied in the setting of 
finite underlying alphabet size $k$, and large block lengths $n$, however 
in numerous applications the underlying natural alphabet best describing
the data might be large. For example, a text document only contains a
small fraction of all the words in the dictionary, and a natural image 
contains only a small portion of all possible pixel values. 
An extreme case of this is when $k=\infty$ and $n=1$ in the 
equation above, showing that $\ered(\cD_\infty)=\infty$~\cite{Kie78,Dav73}. 
The class of all \iid distributions over $\integersp$ is
therefore \emph{too large} to provide meaningful compression schemes
for all distributions. 
These impossibility results led researchers to consider natural 
sub-classes of all \iid distributions and then compress sequences
generated from distributions in these or to consider all \iid
distributions but decompose sequences into natural components
and compress each part separately. 

In recent years, three approaches have been proposed to address
compression over large alphabets.~\cite{OSZ03} considered separate compression of
the \emph{dictionary} that describes the symbols appearing,
and the \emph{pattern} that specifies their order.
For example, the word ``paper'' has pattern 12134 and dictionary 1$\to$p,
2$\to$a, 3$\to$e, 4$\to$r. 
They showed that while the pattern contains almost all of the
sequence entropy, it can be compressed with sublinear worst-case
redundancy, regardless of the alphabet size, hence is \emph{universally
compressible}.

\cite{FSW02} considered the subclass of monotone distributions over
$\integersp$. They showed that even this class is not universally 
compressible, but designed universal codes for all monotone
distributions with finite entropy.
\cite{Sha13} studied the class $\monkn$ of length$-n$ sequences from
monotone distributions over $[k]$, and tightly characterized the
redundancy for any $k=O(n)$. 
Recently,~\cite{AcharyaJOS14:monotone} studied $\monkn$ for much
larger range of $k$ and
in particular showed that the class is
universally compressible for all $k=\exp(o(n/\log n))$
and is not universally compressible for any $k=\exp(\Omega(n))$. 

A third approach was proposed in~\cite{BGG09}. They studied
compression of \emph{envelope classes}, where the probabilities
are bounded by an envelope.
They provide general bounds on the redundancy of envelope classes.
They were motivated by the previous negative results on compressing
\iid distributions over infinite alphabets, and therefore wanted to
consider  classes for which it is possible to design universal codes. 
The upper bounds on the worst-case redundancy are obtained by bounding 
the Shtarkov sum. They provide bounds on the more stringent (for lower 
bounds) average case redundancy by employing 
the redundancy-capacity theorem. 

%

In the next section we introduce and motivate envelope classes and 
describe some of the known results and our new results.

\ignore{\tcr{ REMOVE THIS WHOLE THING??? 
{\color{blue} Remove the next part}
\section*{Put someplace}
\subsection*{Where?}
Coding length-$n$ sequences over $\cX$ equivalently is encoding
symbols from $\cX = \cXn$. Such block compression can be 
treated as a one-shot coding when we treat each  
$\xn\in\cXn$ as a symbol $\symb$. 
As before we consider the infimum over all distributions (not only \iid)
 $\distqn$ over $\cXn$ to define
\begin{align*}
\ered(\dclassn) &\ed \inf_{\distqn}\sup_{\distPn}
D\left(\distPn||Q_n\right),\\
\wrpn &\ed \inf_{\distqn}\sup_{\distPn}\sup_{\xn\in\cXn}
\log \frac{\distPn(\xn)}{Q_n(\xn)},
\end{align*}
as the average and worst case redundancy of $\dclassn$. 
$\dclass$ is \emph{universal} if $R(\dclass) = o(n)$,
\ie redundancy is sublinear in the block-length.
For $\xn\in\cXn$, let 
\[
mlpxn \ed \sup_{\distP\in\dclass}{\distP^n(\xn)},
\]
be the maximum likelihood (ML) probability of $\xn$ and 
$\mlpn$ is the ML distribution.
Since we are compressing 
length-$\lng$ sequences,
\[
\shtarkov(\dclassn)=\sum_{\vecx\in\cXn}mlpxn.
\]}}

\section{Envelope Class: Known Results}

A function $f:\integersp\to\reals^{\ge 0}$ is called an \emph{envelope}.
We abbreviate $f_i\ed f(i)$.
Any envelope $f$ determines an \emph{envelope class}
\[
\Envcls
\ed
\Sets{(p_1,p_2,\ldots)\in\cDinf: p_i\le f_i}.
\]
When $f$ is defined explicitly, we will sometimes denote
$\Envcls$ by $\Envclsi$. For example, one of our main applications
is for the power envelope defined by  $f_i=c\cdot i^{-\alpha}$ and the corresponding
power class is denoted by $\Pwrcalp$.

Envelope classes naturally generalize $\cDk$ for large 
and potentially infinite $k$.
They can incorporate prior distribution knowledge
that can be expressed as an upper bound on the probabilities.

\cite{BGG09} introduced envelope classes and 
proved several results about their redundancy for general, power-law,
and exponential envelopes. 
They called an envelope function $f$ is \emph{summable} if 
$
\sum_{i=1}^\infty f_i<\infty
$,
and used the Shtarkov sum to show that $\wred(\Envcls)<\infty$
if and only if $f$ is summable.
Letting
\[
\barF
\ed
\sum_{i=u+1}^\infty f_{i}
\]
be the tail sum, they showed that
\[
\wred(\Envclsn)
\le
\inf_{u\le n}\Big[n\barF+\frac{u-1}2\log n\Big]+2.
\]
They also used the redundancy-capacity theorem~\cite[Chap 13]{CT91} to
lower bound the expected and therefore worst-case redundancy, but their
expressions are more involved and we refer the reader to their paper. 

They applied these bounds to two important and natural envelope classes.
For the \emph{power-law} envelope class defined by the envelope
$\const\cdot i^{-\alpha}$ for $c>0$ and $\alpha>1$, they showed%
\footnote{$g(n)=O(f(n))$ if $\exists C, N: \forall n\ge N 
  g(n)\le Cf(n)$.}
\begin{equation}
\label{eqn:power_previous}
C_{c,\alpha}\cdot n^{\frac1\alpha}
\le
\ered(\Pwrcalpn)
\le
\wred(\Pwrcalpn)
\le
\Big(\frac{2\const n}{\alpha-1}\Big)^{\frac1\alpha}(\log
n)^{1-\frac1\alpha} + O(1),
\end{equation}
where the constant $C_{c,\alpha}$ depends on $\const$ and $\alpha$.
They also noted that the lower-bound and the upper-bound is order
$(\log n)^{1-\frac1\alpha}$ apart and asked whether one of them is tight.
One of our results shows that the lower bound is tight
up to a constant factor.

For the \emph{exponential-law} envelope class, defined by the envelope
$f_i=\const\cdot e^{-\alpha i}$ where $\const,\alpha>0$ 
they proved that%
\footnote{$g(n)=o(f(n))$ if $\lim_{n\to\infty} \frac{g(n)}{f(n)}=0.$}
\[
\frac{\log^2 n}{8\alpha\log e}(1+o(1))
\le
\wred(\Expcalpn)
\le
\frac{\log^2 n}{2\alpha\log e}+O(1).
\]

\cite{B10} improved these bounds and determined the redundancy up to
the first order term and showed
\begin{align}
\label{eqn:bontemp_exponential}
\wred(\Expcalpn)= \frac{\log^2 n}{4\alpha\log e}+O(\log n\log \log n).
\end{align}

More recently,~\cite{BontempsBG12} extended the arguments of~\cite{B10}
to find tight universal codes for the larger class of
sub-exponential envelope class whose decay is strictly faster
than power-law, but slower than exponential classes.
However, as mentioned in these papers, a tight bound on the redundancy
of heavy-tail envelopes such as power-laws remained elusive. Note
that one of the important goals of~\cite{BontempsBG12} was to obtain
sequential algorithms achieving the redundancy of these classes, not
simply obtaining the bounds as we consider.~\cite{BoucheronGO14}
design efficient codes that are \emph{adaptive} for classes of
envelopes, namely they provide efficient compression schemes for
families of envelopes, without the knowledge of the precise envelope 
being followed.

\section{New Results and Techniques}

Poisson sampling provides simplified analysis in various
machine learning and statistical problems~\cite{MU05}. To the 
best of our knowledge, it was first used for universal compression in~\cite{ADO12},
and~\cite{AcharyaDJOS13} to prove optimal bounds on pattern
redundancy. Recently,~\cite{YangB13}
also apply Poisson sampling to provide simpler coding schemes
for \iid distributions over large alphabets.
They observed that Poisson sampling renders the multiplicities of
each symbol independent and hence can be independently coded.
Using the tilting method they constructed
compression schemes that are optimal upto an additional factor of
$\frac{1}{2}\log\frac nk$ for
$\cDkn$. In a subsequent paper, they showed that it can be extended to sources with memory.
The key difference between~\cite{YangB13} and this paper is that
the former focuses on finding efficient coding schemes for $\cDkn$ and sources with memory, while
we use the Poisson model as a proof technique to simplify the
computation of the
redundancy of \iid distributions and various envelope classes.

In section~\ref{sec:poisson_model} we describe the Poisson
sampling model. We then relate the redundancy of this model 
to the fixed length model. In
Section~\ref{sec:single_poisson} we consider the class of 
simple Poisson distributions with bounded means. 
In Theorem~\ref{thm:env_class}, we give \emph{single-letter}
bounds on the worst-case redundancy
of general envelope classes in terms of redundancy of single Poisson
classes. The bounds are simply summations
of such Poisson redundancies. The upper and lower bounds in this theorem
differ by an additive factor of $O(\log n)$. 

We first apply the results to the class $\cDkn$ and present a
short argument that bounds its redundancy tightly up to the
first order term.
We then apply these bounds to find the worst case redundancy
of power-law class. Even though we give a single letter bound that is
tight barring additive logarithmic factors, we now focus on providing
\emph{closed} form bounds that are tight to a multiplicative factor of
4. In particular, this strengthens Equation~\eqref{eqn:power_previous} 
answering a question posed in~\cite{BGG09}.
This also shows that in fact the lower bound
of~\cite{BGG09} on the average redundancy is within a constant 
factor from the worst case.
We then bound the redundancy of exponential classes improving the 
second order term in Equation~\eqref{eqn:bontemp_exponential}.
\section{Preliminary Results}
We first mention several simple redundancy results
that will be useful later in the paper.

\subsection{The Shtarkov Sum}
Worst-case redundancy can be easily expressed as a sum of probabilities.
Let $\dclass$ be a collection of distributions over $\cX$.
The \emph{maximum likelihood} probability of $x\in\cX$ is
\[
\mlpx
\ed 
\sup_{\distP\in\dclass} \distP(\symb),
\] 
the highest probability assigned to it by any distribution in $\dclass$.
The \emph{maximum-likelihood-}, or \emph{Shtarkov sum}~\cite{Sht87} is 
\[
\shtarkov(\dclass)
\ed
\sum_{\symb\in\cX} \mlpx.
\]
It is easy to see that
\[
\wred(\dclass)
=
\log \left(\shtarkov(\dclass)\right).
\]
This simple formulation allows for simple proofs of several
results for worst-case redundancy.

\subsection{Basic Redundancy Properties}

All the results mentioned in this section apply to both
worst-case and expected redundancy, but for simplicity
we present the proof for the worst-case.
Unless otherwise mentioned, the following assumes that the distributions
are over a set $\cX$. 

\begin{Lemma}[Subset redundancy]
If $\dclass'\subseteq\dclass$, then
\[
\wred(\dclass')
\le
\wred(\dclass).
\]
\end{Lemma}
\begin{proof}
By Shtarkov's Sum,
\[
\shtarkov(\dclass')
=
\sum_{x\in\cX}\max_{P\in\dclass'} P(x)
\le
\sum_{x\in\cX}\max_{P\in\dclass} P(x)
=
\shtarkov(\dclass).\qedhere
\]
\end{proof}

\begin{Lemma}[Union redundancy]
For all distribution collections $\dclass_1\upto\dclass_c$,
\[
\max_{1\le i\le c}\wred(\dclass_i)
\le
\wred(\union_{i=1}^c\dclass_i)
\le
\max_{1\le i\le c}\wred(\dclass_i)+\log c.
\]
\end{Lemma}
\begin{proof}
The lower bound follows from subset redundancy.
For the upper bound, let $\cX_i\subseteq\cX$ be the set of $x$'s
that are assigned the highest probability by a distribution in $\cP_i$
and, in case of ties, not in any $\distP_j$ for $j<i$.
Then,
\[
\shtarkov(\union_{i=1}^c\dclass_i)
=
\sum_{i=1}^c \sum_{x\in\cX_i} \max_{P\in \dclass_i} P(x)
\le
\sum_{i=1}^c \sum_{x\in\cX} \max_{P\in \dclass_i} P(x)
=
\sum_{i=1}^c \shtarkov(\dclass_i)
\le
c\cdot\max_{1\le i\le c}\shtarkov(\dclass_i).\qedhere
\]
\ignore{The upper bound follows from the definition of Shtarkov's sum and
the fact that maximum over a union of sets is lesser than that
sum of maximums within each set.
\[
\shtarkov(\union_{i=1}^c\dclass_i)
=
\sum_{x \in \cX}  \max_{P\in \union_{i=1}^c\dclass_i} P(x)
\leq 
\sum_{x \in \cX} \sum^c_{i=1} \max_{P\in \dclass_i} P(x)
= 
\sum^c_{i=1}  \sum_{x \in \cX}  \max_{P\in \dclass_i} P(x)
= 
\sum^c_{i=1} \shtarkov(\dclass_i)
\leq 
c\cdot\max_{1\le i\le c}\shtarkov(\dclass_i).\qedhere
\]}
\end{proof}

If $\distP$ is a distribution over $\cX$ and $f:\cX\to\cY$, then
$f(\distP)$ is the distribution over $\cY$ defined by
$[f(\distP)](y)=\distP(f^{-1}(y))$.
Similarly, if $\dclass$ is a collection of distributions over $\cX$,
then $f(\dclass)\ed\sets{f(\distP):\distP\in\dclass}$.
\begin{Lemma}[Function redundancy]
$\wred(f(\dclass))\le\wred(\dclass)$ with equality iff for every $y\in\cY$,
all $x\in f^{-1}(y)$ are assigned their highest probability
by the same distribution in $\dclass$.
\end{Lemma}
\begin{proof}
Follows from the maximum-likelihood sum, with equality under
the above condition,
\[
\shtarkov(f(\dclass))
=
\sum_{y\in\cY}\max_{\distP\in\dclass}\sum_{x\in f^{-1}(y)}\distP(x)
\le
\sum_{y\in\cY}\sum_{x\in f^{-1}(y)}\max_{\distP\in\dclass}\distP(x)
=
\sum_{x\in\cX}\max_{\distP\in\dclass}\distP(x)
=
\shtarkov(\dclass).\qedhere
\]
\end{proof}

\begin{Corollary}[Bijection redundancy]
\label{cor:bijection_redundancy}
If $f$ is 1-1, then
\[
\wred(f(\dclass))=\wred(\dclass).\eqed
\]
\end{Corollary}

If $\distP_\cX$ and $\distP_\cY$ are distributions over $\cX$ and
$\cY$ respectively, then $\distP_\cX\times\distP_\cY$ is the
distribution over $\cX\times\cY$ defined by 
$[\distP_\cX\times\distP_\cY](x,y)=\distP_\cX(x)\cdot\distP_\cY(y)$.
Similarly, if $\dclass_\cX$ and $\dclass_\cY$ are collections of distributions
over $\cX$ and $\cY$ respectively, then 
$\dclass_\cX\times\dclass_\cY
\ed
\sets{
\distP_\cX\times\distP_\cY:
\distP_\cX\in\dclass_\cX,\distP_\cY\in\dclass_\cY}.
$
\begin{Lemma}[Product redundancy]
\label{lem:product_redundancy}
For all $\dclass_\cX$ and $\dclass_\cY$,
\[
\wred(\dclass_\cX\times\dclass_\cY)
=
\wred(\dclass_\cX)+\wred(\dclass_\cY).
\]
\end{Lemma}
\begin{proof}
By the Shtarkov sum,
\[
\shtarkov(\dclass_\cX\times\dclass_\cY)
=
\sum_{(x,y)\in\cX\times\cY}\mlp(x,y)
=
\sum_{(x,y)\in\cX\times\cY}\mlp(x)\cdot\mlp(y)
=
\Paren{\sum_{x\in\cX}\mlp(x)}
\cdot
\Paren{\sum_{y\in\cY}\mlp(y)}
=
\shtarkov(\dclass_\cX)\cdot\shtarkov(\dclass_\cY).\qedhere
\]
\end{proof}
For class $\dclass$ over $\cX\times \cY$ consisting of 
product distributions with $\dclass_\cX$ and $\dclass_\cY$ being the
collection of all marginal distributions over $\cX$ and let $\cY$, respectively.
\ignore{For example, 
if
$\cX=\sets{x,x'}$ and
$\cY=\sets{y,y'}$,
and
\[
\dclass
=
\sets{(p_{x,y},p_{x,y'},p_{x',y},p_{x',y'}):p_{x,y}=p_{x',y'}, p_{x,y'}=0},
\]
then $\dclass_\cX=\sets{(p_x,p_{x'}):p_x\le p_{x'}}$ and 
$\dclass_\cY=\sets{(p_y,p_{y'}):p_y\ge p_{y'}}$.}
Then, $\dclass\subseteq\dclass_\cX\times\dclass_\cY$, and
by combining product and subset redundancy,
\begin{Corollary}[Marginal redundancy]
\label{cor:marginal_redundancy}
For every product distribution class $\dclass$ over $\cX\times \cY$,
\[
\wred(\dclass)\le\wred(\dclass_\cX)+\wred(\dclass_\cY).\eqed
\]
\end{Corollary}
\begin{Remark}
The result can be generalized to product distributions over a
countable number of coordinates $\cX_1\times\cX_2\times\ldots$.
\end{Remark}
We now prove some results for the redundancy of \iid distributions. 

\subsection{Implications to \iid Distributions}
Recall that if $\distP$ is a distribution over $\cX$, 
then $\distP^n$ is the induced distribution over $\cX^n$, the sequences of length $n$.
\ignore{Suppose $\dclass$ is a collection of product (independent)  
distributions over $\cX\times \cY$, \ie
each element in $\dclass$ is a distribution of the 
form $\distP_1\times \distP_2$, where $\distP_1$ and $\distP_2$ are
distributions over $\cX$ and $\cY$ respectively. 
Let $\dclass_{\cX}$ and $\dclass_{\cY}$ be the class of
marginals over $\cX$ and $\cY$. 
Then the redundancy of $\dclass$ is at most the sum of the marginal
redundancies.
\begin{Lemma}[Redundancy of products~\cite{AcharyaDJOS13}]
\label{lem:product_redundancy}
For a collection $\dclass$ of product distributions over
$\cX\times\cY$, 
\begin{align*}
\wred(\dclass) \leq \wred(\dclass_\cX) + \wred(\dclass_\cY).
\end{align*}
Furthermore, if $\dclass = \dclass_\cX \times \dclass_{\cY}$ then 
equality holds. 
\end{Lemma} 
\begin{proof}
For any $(\symb, \symby)\in\cX\times \cY$, 
\[
\sup_{(\distP_1,\distP_2)\in\dclass}\distP_1(\symb)\distP_1(\symby)\le 
\sup_{\distP_1\in\dclass_{\cX}}\distP_1(\symb)\sup_{\distP_2\in\dclass_{\cY}}\distP_2(\symby).
\]
Therefore,
\begin{align*}
\shtarkov (\dclass) =& \sum_{(\symb,\symby)\in\cX\times\cY}\sup_{(\distP_1,\distP_2)\in\dclass}\distP_1(\symb)\distP_2(\symby) \\
\le& \sum_{a\in\cX}\sup_{\distP_1\in\dclass_\cX}\distP_1(\symb)
  \sum_{b\in\cY}\sup_{\distP_2\in\dclass_\cY}\distP_2(\symby)\\
 =& \shtarkov (\dclass_\cX) \shtarkov (\dclass_\cY),
\end{align*}
and the lemma  
follows by taking logarithms.
When all possible combinations of marginals is possible, then 
the inequality becomes an equality.
\end{proof}}

\begin{Lemma}
\label{lem:properties_redundancy}
\begin{enumerate}
\item
\emph{Monotonicity:} For all $\lng$, $\wred(\dclass^{\lng+1})\ge\wred(\dclassn)$.
\item
\emph{Subadditivity:} For any $n_1$, $n_2$ 
$\wred(\dclass^{n_1+n_2})\le \wred(\dclass^{n_1})+\wred(\dclass^{n_2})$.
\end{enumerate}
\end{Lemma}
\begin{proof}
Monotonicity follows by marginalizing the $(n+1)$th coordinate
and taking logarithm in the following.
\begin{align*}
\shtarkov(\dclass^{n+1}) &= \sum_{x_1^{n+1}\in\cX^{n+1}}\sup_{\distP\in\dclass}\distPn(x_1^{n+1})\\
&\ge \sum_{\xn\in\cXn}\sup_{\distP\in\dclass}\left[\distPn(\xn)\left(\sum_{x_{n+1}\in\cX}\distP(x_{n+1})\right)\right]\\
&=\sum_{\xn\in\cXn}\sup_{\distP\in\dclass}\distPn(\xn)\\
&=\shtarkov(\dclassn).
\end{align*}
For subadditivity, we give the proof of~\cite{BGG09}.
\begin{align*}
\shtarkov(\dclass^{n_1+n_2}) &=
 \sum_{x_1^{n_1+n_2}\in\cX^{n_1+n_2}}\sup_{\distP\in\dclass}\distP(x_1^{n_1+n_2})\\
&=
\sum_{x_1^{n_1+n_2}\in\cX^{n_1+n_2}}\sup_{\distP\in\dclass}\Big[\distP(x_1^{n_1})\distP(x_{n_1+1}^{n_2})\Big]\\
&\le
\sum_{x_1^{n_1+n_2}\in\cX^{n_1+n_2}}\sup_{\distP\in\dclass}\distP(x_1^{n_1})
\sup_{\distP\in\dclass}\distP(x_{n_1+1}^{n_2})\\
&= \Big[\sum_{x_1^{n_1}\in\cX^{n_1}}\sup_{\distP\in\dclass}\distP(x_1^{n_1})\Big]\cdot
\Big[\sum_{x_1^{n_2}\in\cX^{n_2}}\sup_{\distP\in\dclass}\distP(x_{1}^{n_2})
\Big]\\
&=\shtarkov(\dclass^{n_1})\cdot \shtarkov(\dclass^{n_2}).
\end{align*}
Taking logarithm proves the second part. 
\end{proof}

\subsection{Type Redundancy}
The \emph{multiplicity} $\mlt_{\xn}(x)$ is the number of times
a symbol $x$ appears in a sequence $\xn$.
The \emph{type} of a sequence $\xn$ over $[k]$ is the $k$-tuple 
\[
\type(\xn)
\ed
\Paren{\mlt_{\xn}(1),\mlt_{\xn}(2)\upto\mlt_{\xn}(k)}
\]
of multiplicities of the $k$ symbols in the sequence
$\xn$, see \eg~\cite{CT91}. 
For example, for $k=4$, $\type(11313)=(3,0,2,0)$.
Note that a type is a $k$-tuple of non-negative integers summing to $n$
and that for $k=\infty$, the type has infinite length.

Let $\type(\distPn)$ be the distribution induced by $\distPn$ over types.
It is easy to see that $\type(\distPn)$ is the multinomial
(specifically, $k$-\emph{nomial}) distribution with parameters
$n$ and $(p_1\upto p_k)$, namely
\[
\distP\Paren{\type(\Xn)=(\mlt_1\upto\mlt_k)}
=
\binomial{n}{\mlt_1\upto\mlt_k}\cdot\prod_{i=1}^k p_i^{\mlt_i}.
\]
Note that this definition applies also for infinite $k$ as at most
$n$ multiplicities are non-zero.
We let
\[
\type(\cP^n)=\{\type(\distPn):\distP\in\dclass\}
\]
be the set all distributions induced by $\distPn$ over types,
namely the set of all $k$-nomial distributions whose first parameter
is $n$.

It is well known that for \iid distributions,
the redundancy of sequences
is equal to the redundancy of types, which follows since 
any \iid distribution assigns
the same probability to all sequences of the same type.
\begin{Lemma}
\label{lem:type_redundancy}
For $\red\in\{\ered,\wred\}$, 
\[
\red(\type(\cP^n))=\red(\cP^n).
\]
\end{Lemma}
\begin{proof}
We prove the result for worst case since it has simpler expressions. 
The average case follows similarly from the definition of average 
redundancy and convexity of KL divergence. 
Any \iid distribution assigns the same probability to all sequences with
the same type. Therefore, such sequences have the same maximum 
likelihood probability. We show that the Shtarkov sums of the two classes
are the same and hence they have same redundancy.
\[
\shtarkov(\cP^n) = \sum_{\xn\in\cX^n}\mlpxn
= \sum_{\type}\sum_{\xn:\type(\xn)=\type} \hat{P}^n(\xn)
= \sum_{\type}\hat{P}^n(\type)
= \shtarkov(\type(\cP^n)).\qedhere\]
\end{proof}

\section{The Poisson Model}
\label{sec:poisson_model}
Recall that if $\distP$ is a distribution over $\cX$, then
$\distPn$ is the distribution over $\cXn$ derived by
sampling according to $\distP$ independently $n$ times.
A significant difficulty in analyzing such distributions is that
although the samples are chosen independently, the number
of appearances of different symbols are dependent.
For example, they always add to $n$.

To overcome this difficulty, the Poisson model also considers
independent samples according to $\cP$, but eliminates the
dependence between symbols by generating not a fixed, but a
variable number of samples that follow a Poisson distribution.

We first recall the Poisson distribution and one of its concentration
result, then define Poisson sampling and mention some of its
properties, and finally relate its redundancy to fixed-length
redundancy.

\subsection{The Poisson Distribution}
The Poisson distribution $\Psnlmb$ with parameter $\lambda$ assigns to
$\psnarg\in\naturals$ probability 
\[
\Psnlmbarg\ed e^{-\lambda}\frac{\lambda^{\psnarg}}{\psnarg!}.
\]
The Poisson distribution concentrates exponentially around its mean, helping
relate the redundancy of Poisson- and fixed-length sampling.
\begin{Lemma}(\cite{MU05})
\label{lem:poisson_tail_bound}
Let $\varx\sim\Psnlmb$, then for $x\ge\param$,
\[
\Psnlmb(X \geq x) \leq \exp\left(-\frac{(x-\param)^2}{2x}\right),
\]
and for $x\le\param$,
\[
\Psnlmb(X \leq x) \leq \exp\left(-\frac{(x-\param)^2}{2\lambda}\right).
\eqed
\]
\end{Lemma}

\subsection{Poisson Sampling and its Properties}
Poisson-length sampling replaces a fixed number $n$ of samples
by a random number $\Np\sim\Psnn$ of samples, and then 
sampling $P$ independently $\Np$ times.
Consequently, the distribution is over the set $\cX^*\ed\cup_{i=0}^\infty\cX^i$
of finite strings over $\cX$, where $\cX^0$ contains just the empty string.
Therefore the probability of a sequence $\xnp\in\cX^*$ is 
\[
\distPPsnn(\xnp)
=
\Psnnnp\cdot\distP^{\np}(\xnp)
=
\Psnnnp\cdot\prod_{i=1}^{\np} P(x_i).
\]
The following lemma, which 
along with its elementary proof can be found in~\cite{MU05},
states that conditioned on $\Np=\np$, the distribution 
that $\distPPsnn$ induces on sequences is identical to that
of $\distPnp$, and that under Poisson sampling, the multiplicity $\MltPsnn_i$
of symbol $i$ is distributed Poisson, and that multiplicities of different symbols
are independent. 
\begin{Lemma}
\label{lem:poisson_independent}
For all $k$, $\distP\in\cDk$, and $n$,
for $1\le i\le k$, $\MltPsnn_i\sim\Psns{np_i}$ independently of each other.
\hfill\qed
\end{Lemma}
Therefore, for a distribution 
$\distP=(p_1, p_1, \ldots)$ over $\cX=\{1,2,\ldots\}$
\begin{align}
\label{eqn:poi_type}
\type(\cP^{\Psnn})=(\Psns{np_1},\Psns{np_2},\ldots),
\end{align}
where each coordinate is an independent Poisson distribution. 

\begin{Lemma}
\label{lem:poisson_conditioned}
For all $k$, $\distP\in\cDk$, and $n$,
for all $\np$,
$\distPPsnn(\xnp|\Np=\np)=\distPnp(\xnp)$.\hfill\qed
\end{Lemma}

Similar to $\dclassn$, let 
\[
\dclasspsnn \ed \{\distPPsnn:\distP\in\dclass\}
\]
be the class of distributions over $\cX^*$, where each distribution
consists of a distribution $\distP\in\cP$ sampled independently
a random $\Psnn$ times. 

The next lemma relates Shtarkov sums for Poisson and fixed-length
sampling.
\begin{Lemma}
\label{lem:shtarkov_poisson}
For every distribution class $\dclass$, 
\[
\shtarkov\Paren{\dclasspsnn}
=
\sum_{n'=0}^\infty\Psnss{n}{n'}\cdot\shtarkov\left(\dclass^{n'}\right).
\]
\end{Lemma}
\begin{proof}
By Lemma~\ref{lem:poisson_conditioned},
the Shtarkov sum conditioned on the length 
does not change, namely for a sequence $\xnp$
the same distribution attains maximum likelihood
for both $\Psnn$ sampling and sampling \iid $n'$ times.
Therefore,
\begin{align*}
\shtarkov\left(\dclasspsnn\right) = \sum_{n'\ge0}
\Psnnnp\sum_{\xnp}\sup_{\distP\in\dclass} \distP^{n'}(\xnp)
= \sum_{n'\ge0}
\Psnnnp\shtarkov\left(\dclass^{n'}\right),
\end{align*}
where in the first step we sum maximum likelihoods of 
sequences by their lengths.
\end{proof}

We would like to use the independence of multiplicities to obtain
simpler bounds on the redundancy of \iid distributions. Toward
this, we first show in the next result that under mild assumptions, 
the redundancy of a class under Poisson sampling is \emph{close}
to the redundancy under fixed length sampling. The bound we are more
interested in, namely proving upper bounds on $\wrpn$ holds 
even without these assumptions, namely for arbitrary classes. 
\begin{Theorem}
\label{thm:poisson_fixed}
For any class $\dclass$ 
\[
\wred(\dclassn)\le \wred(\dclasspsnn)+1.
\]
Furthermore, for $n \geq 4$ if 
 $\wred(\dclassn)<n/16$ and   $\nlower\ed \lfloor n-3\sqrt{n\wred(\cP^{n}) }\rfloor$, then
\[
\wred(\dclass^{\Psns\nlower})
\le
\wred(\dclassn).
\]
\end{Theorem}
\begin{Remark}
\label{rem:assumptions_redundancy}
The first part of the theorem in some
cases can be used to verify if the conditions for the 
second part hold. This is a potential method of obtaining
lower bounds given good upper bounds.
\end{Remark}

\begin{proof}[Proof of Theorem~\ref{thm:poisson_fixed}]
By Lemma~\ref{lem:shtarkov_poisson}, 
\begin{align*}
\shtarkov (\cP^{\Psnn}) &\stackrel{(a)}{=} \sum_{n'\ge0}\Psnnnp \shtarkov (\cP^{n'})\\
&\stackrel{(a)}{\ge} S(\cP^n)\sum_{n'\ge n}\Psnnnp\\
&\stackrel{(b)}{\ge} \frac12  \shtarkov (\cP^n).
\end{align*}
where
$(a)$ from monotonicity and $(b)$ from the fact that
if the mean is an integer, then
the 
median of a Poisson distribution is larger than its mean.
Taking logarithms proves the first part. 

Let $\nlower\ed \lfloor n-3\sqrt{n\wred(\cP^{n}) }\rfloor$, then by monotonicity
and the fact that median is larger than the mean, 
\begin{align*}
\shtarkov(\cP^{\Psns\nlower})
=
\sum_{\np}\Psnss\nlower\np\cdot\shtarkov(\cP^{\np})
\le
\frac12\shtarkov(\dclassn)+\sum_{\np\ge n}\Psnss\nlower\np\shtarkov(\cP^{\np}).
\end{align*}
By the subadditivity 
and monotonicity, 
\[
\wred(\dclass^{\np})\le\wred(\dclass^{\left \lceil \frac{n'}{n} \right\rceil n}) \leq \left\lceil \frac{n'}{n} \right\rceil
\wred(\dclassn) \leq \left( \frac{n'}{n} +1 \right) \wred(\dclassn),
\]
and hence $\shtarkov(\cP^{\np}) \leq \shtarkov (\cP^n) (\shtarkov (\cP^n) )^{ \frac{n'}{n}}$.
Therefore,
\begin{align*}
 \sum_{\np\ge n}\Psnss\nlower\np\shtarkov(\cP^{\np})
 & \leq 
  \shtarkov (\cP^n)  \sum_{\np\ge n}\Psnss\nlower\np (\shtarkov (\cP^n) )^{ \frac{n'}{n}}\\
  & =  \shtarkov (\cP^n) \sum_{\np\ge n} \frac{1}{\np!}e^{-\nlower} \nlower^{n'} (\shtarkov (\cP^n) )^{ \frac{n'}{n}} \\
  & = \shtarkov (\cP^n)\exp(\nlower(\shtarkov (\cP^n)^{\frac{1}{n}} -1))\sum_{\np\ge n}\Psnss{\nlower\shtarkov (\cP^n)^{\frac{1}{n}}}\np \\
  & \stackrel{(a)}{\leq} \shtarkov (\cP^n)\exp\left[\nlower( \shtarkov (\cP^n)^{\frac{1}{n}} -1)
  -\frac{(\nlower\shtarkov (\cP^n)^{\frac{1}{n}}-n)^2}{2n}  \right] \\
  & \stackrel{(b)}{\leq} \frac{1}{2}\shtarkov (\cP^n).
\end{align*}
$(a)$ uses Poisson tail bounds and  $\nlower (\shtarkov (\cP^n)^{\frac{1}{n}}) \leq n$.
$(b)$ follows from the bounds on $\wred(\cP^n)$ and $\nlower$.
\end{proof}

We finally show that sequence redundancy is equal to type redundancy
under Poisson sampling. 
\begin{Lemma}
\label{lem:type_redundancy}
For $\red\in\{\ered,\wred\}$, 
\[
\red(\type(\cP^{\Psnn}))=\red(\cP^{\Psnn}).
\]
\end{Lemma}
\begin{proof} We again show only the worst case.
\[
\shtarkov(\cP^{\Psnn})\stackrel{(a)}{=}
\sum_{n'=0}^\infty\Psnss{n}{n'}\cdot\shtarkov\left(\dclass^{n'}\right)
\stackrel{(b)}{=}
\sum_{n'=0}^\infty\Psnss{n}{n'}\cdot\shtarkov\left(\type\left(\dclass^{n'}\right)\right)
= \red(\type(\cP^{\Psnn})),
\]
where $(a)$ uses Lemma~\ref{lem:shtarkov_poisson} and 
$(b)$ follows from Lemma~\ref{lem:type_redundancy}.
\end{proof}

This lemma in conjunction with Theorem~\ref{thm:poisson_fixed}
and~\ref{thm:env_class}
reduces the problem of bounding the redundancy of types under 
Poisson sampling. Under Poisson sampling types of sequences 
are tuples of independent Poisson random variables, with distribution
given by Equation~\eqref{eqn:poi_type}. 

In the next section we study the simple class of Poisson
distributions with bounded means. We use this class as a primitive to
bound the redundancy of envelope classes in Section~\ref{sec:env_class}.

\section{The Redundancy of Bounded Poisson Distributions}
\label{sec:single_poisson}
Let 
\[
\cPPsnupr
\ed
\{
\Psnlmb: \lambda\le\upper\}
\]
be the class of all Poisson distributions with mean $\le \upper$.
We approximate the class's redundancy, and will later apply it
to approximate the redundancy of all envelope classes.

\begin{Lemma}
\label{lem:one_poisson}
For $\upper\le 1$,
\[
\wred\Big(\cPPsnupr\Big)=\log \left(2-e^{-\upper}\right)\le \upper\log
e,
\]
and for $\upper>1$,
\[
\log \sqrt{\frac{2\upper+2}{\pi}}
\le
\wred\Paren{\cPPsnupr}
\le
\log\Paren{\sqrt{\frac{2\upper}\pi}+2}.
\]
\end{Lemma}
\begin{proof}
Of all Poisson distributions, the probability of $i\in\naturals$
is maximized by $\Psni$. 
Hence of all distributions in $\cPPsnupr$,
the probability of $i\in\naturals$ is maximized by the
Poisson distribution with mean
\[
\argmax_{\lambda\le\upper}\Psnlmbi
=
\begin{cases}
i & \text{for } i\le \upper, \\
\upper & \text{for } i\ge \upper. \end{cases} 
\]
The Shtarkov sum is therefore 
\[
S\Paren{\cPPsnupr}
=
\sum_{i=0}^{\floor{\upper}}\psnfrm{i}{i} +
\sum_{i=\floor{\upper}+1}^{\infty}\psnfrm{\upper}{i}.
\]

For $\upper\le1$,
\[
S\Big(\cPPsnupr\Big)
=
1+\sum_{i=1}^\infty\psnfrm{\upper}i
=
1+ (1-e^{-\upper})
=
2-e^{-\upper}
\le
e^\upper,
\]
where the middle equality follows from
$\sum_{i=0}^\infty\psnfrm{\upper}i=1$,
and the inequality from the arithmetic-geometric inequality.

For $\upper>1$, first observe that Stirling's approximation, stating
that for any $n$, there is some
$\theta_n\in(\frac1{12n+1},\frac1{12n})$
such that
\begin{align}
\label{eqn:stirling}
n!
=
\sqrt{2\pi n}\left(\frac n e \right)^n e^{\theta_n},
\end{align}
implies that for $i\ge1$
\[
\psnfrm ii
\le
e^{-i}\frac{i^i}{\sqrt{2\pi i}(\frac{i}{e})^i}
\le
\frac1{\sqrt{2\pi i}}, 
\]
and using $e^{-x}\ge 1-x$, 
\[
\psnfrm ii
\ge
e^{-i}\frac{i^i}{\sqrt{2\pi i}e^{\frac1{12i}}(\frac{i}{e})^i}
=
\frac{e^{-\frac1{12i}}}{\sqrt{2\pi i}}
\ge\frac 1 {\sqrt{2\pi i}} -\frac 1{12\sqrt{2\pi}}\frac1{i^{3/2}},
\]

Hence,
\[
\shtarkov\Big(\cPPsnupr\Big)
<
2+\sum_{i=1}^{\floor{\upper}}e^{-i}\frac{i^i}{i!}
\le
2+\sum_{i=1}^{\floor{\upper}}\frac1{\sqrt{2\pi i}}
\le
2 + \sqrt{\frac{2\upper}{\pi}},
\]
where the last inequality follows from 
a simple integration.

For the lower bound, 
\begin{align*}
\shtarkov\Big(\cPPsnupr\Big) 
\ge 1+ \sum_{i=1}^{\floor{\upper}}e^{-i}\frac{i^i}{i!}
\ge 1 + \sum_{i=1}^{\floor{\upper}}\frac 1 {\sqrt{2\pi i}} -\frac 1{12\sqrt{2\pi}}\frac1{i^{3/2}}.
\end{align*}
The lower bound follows by integrating this expression.
\end{proof}

\section{The Redundancy of Envelope Classes}
\label{sec:env_class}
We use the redundancy of bounded-Poisson classes to characterize
that of envelope classes.
Given the similarity in redundancy of fixed-length and Poisson-sampled
distributions, determined in Theorem~\ref{thm:poisson_fixed}, we
consider envelope classes under the easier to analyze Poisson
sampling.

Let $f$ be a summable envelope. When a distribution
$\distP=(p_1,p_2,\ldots)\in\Envcls$ is sampled $\Psnn$ times,
symbol $i\in\integersp$ appears
$\Psns{n p_i}$ times, where $n p_i\le n f_i$.
Let 
\[
\slower_f
\ed 
\min\sets{l:\sum_{i=l}^\infty f_i<1}
\]
be the smallest integer whose tail sum is $<1$, or equivalently
$\sum_{i=\slower_f}^\infty n f_i<n$.
Since $f$ is summable, $\slower_f$ is finite.

Our main result for envelope classes provides simple lower and upper
bounds for their redundancy in terms of the redundancy of bounded Poisson
classes. Note that the upper and lower bounds differ by $\slower_f-1$
terms. By Lemma~\ref{lem:one_poisson}, they are tight up to an
additive $\slower_f\cdot\log (2+\sqrt{2n/\pi})=O_f(\log n)$ term,
where $O_f$ implies a constant determined by $f$.

\begin{Theorem}
\label{thm:env_class}
For any envelope $f$ and any $n$,
\[
\sum_{i=\slower_f}^\infty\wred\paren{\cPPsns{n f_i}}
\le
\wred\left(\Envclspsnn\right)
\le
\sum_{i=1}^\infty\wred\paren{\cPPsns{n f_i}}.
\]
\end{Theorem}
\begin{proof}
By Lemma~\ref{lem:type_redundancy}, it suffices to consider the redundancy of 
types of sequences. 
By Equation~\eqref{eqn:poi_type}, the class of type distributions
induced by $\Envcls$ under Poisson sampling is
\[
\type\Paren{\Envcls^{\Psnn}}
=
\prod_{i=1}^\infty \cPPsns{np_i}
\subseteq
\prod_{i=1}^\infty \cPPsns{nf_i}.
\]
By Corollary~\ref{cor:marginal_redundancy} generalized to a
countable number of dimensions
\[
\wred\Paren{\Envclspsnn}
=
\wred\Paren{\type\Paren{\Envclspsnn}}
\le
\sum_{i=1}^\infty \wred\Paren{\cPPsns{nf_i}}.
\]
For the lower bound, note that 
\[
\sum_{i\ge\slower_f}nf_i<n,
\]
and therefore all product distributions in 
\[
\cPPsns{nf_{\slower_{f}}}\times\cPPsns{nf_{\slower_{f}+1}}\times\ldots
\]
are valid projections of a distribution in $\Envclspsnn$ along
the coordinates $i\ge\slower_f$. Applying
Lemma~\ref{lem:product_redundancy} proves the lower bound.
\end{proof}

We now apply this theorem to power law and exponential 
envelope classes. 

%

\section{\iid Sequences over Small Alphabet}

As the first application of Poisson sampling, we study the
redundancy of $\cDkn$ in the range of small $k$ and
$n$ increasing asymptotically.~\cite{SW10}
show that for $k=o(n)$, 
\[
\wred(\cDkn) = \frac {k-1}2\log n -\frac k2 \log k +\frac k2 \log e +o(k).
\]
This is a more refined expression than the one presented in the 
introduction. 

Using the tools developed for Poisson sampling, we derive an
extremely short argument that bounds the redundancy 
of this class up to first order terms.  

For a length$-n$ sequence with type $\type'=\Paren{\mlt_1, \ldots,
  \mlt_{k}}$, let the \emph{abbreviated type} be 
\[
\type'\ed \Paren{\mlt_1\upto\mlt_{k-1}},
\]
the $(k-1)-$tuple by dropping the last multiplicity. 
Then, for length$-n$ sequences, 
there is a bijection between $\type$ and $\type'$.
By Corollary~\ref{cor:bijection_redundancy},
\begin{align*}
\wred(\cDkn) &= \wred(\type(\cDkn))\\
&= \wred(\type'(\cDkn))\\
&\stackrel {(a)}{\le} \wred(\type'(\cDkpsnn))+1\\
&\stackrel {(b)}{\le} (k-1)\wred(\cPPsns{n})+1\\
&\stackrel {(c)}{\le} (k-1)\log \frac{3\sqrt n}\pi +1\\
&\stackrel{(d)}{\le} \frac{k-1}2\log {n} +   k\log\frac 3\pi+o(k), 
\end{align*}
where $(a)$ follows similar to Theorem~\ref{thm:poisson_fixed}, 
$(b)$ uses Lemma~\ref{lem:product_redundancy}, $(c)$ follows from
Lemma~\ref{lem:one_poisson} by using
$2+\sqrt{2n/\pi}<3\sqrt{n}$, 
and $(d)$ by a simple expansion. 

We now prove a lower bound along similar lines. Let
$n'=n-n^{3/4}k^{1/4}\log n$.
\begin{align*}
\wred(\cDkn) &\stackrel{(a)}{\ge} \wred(\cDkpsnnp)\\
&\stackrel{(b)}{\ge}(k-1)\wred(\cPPsns{\frac n{k-1}})\\
&\stackrel{(c)}{\ge} \frac{(k-1)}2\log \frac{2 n'}{\pi (k-1)} \\
&\ge \frac{(k-1)}2\log n -\frac k2\log k - \frac k2\log \frac \pi 2
+o(k), 
\end{align*}
where $(a)$ uses Theorem~\ref{thm:poisson_fixed}, 
$(b)$ follows from the equality condition of
Lemma~\ref{lem:product_redundancy}, 
$(c)$ uses Lemma~\ref{lem:one_poisson}. 

We note that the lower bound is off by $O(k)$ and 
the upper bound by $O(k\log k)$.

\section{Power Law Envelopes}

We apply Theorem~\ref{thm:env_class} to the power law class and provide bounds on 
redundancy that are at most a factor 4 apart. The upper bound improves results of~\cite{BGG09}, 
and proves that their lower bound on the average
redundancy is within a constant factor of the worst
case redundancy.
This resolves an open problem posed in~\cite{BGG09} and is stated in the following 
theorem.
\begin{Theorem}
\label{thm:power_law}
For large $n$ 
\begin{align*}
(\const n)^{1/\alpha}\Big[\frac{\alpha\log e}2+\frac{\log
  e}{2(\alpha-1)}-\frac{\log \frac \pi 2}2\Big](1-o_n(1))\le 
\wred(\Pwrcalpn)\le
(\const n)^{1/\alpha}\Big[\frac {\alpha\log e}2+\frac{\log e}{\alpha-1}+\log 3\Big]+1.
\end{align*}
\end{Theorem}
\begin{proof}
We  bound the redundancy of $\Pwrcalp^{\Psnn}$ and
then apply Theorem~\ref{thm:poisson_fixed} to obtain the result for $\Pwrcalpn$.

\subsubsection{Upper Bound}
For the power-law class $\Pwrcalp$, 
$np_i\le nf_i = \frac{\const n}{i^{\alpha}}$.
Let $\trans\ed(\const n)^{1/\alpha}$, then
\begin{align*}
nf_i \ge 1 \text{ for } i\le \trans \text{ and } nf_i<1
\text{ for } i>\trans.
\end{align*}


\begin{align*}
\wred(\Pwrcalp^{\Psnn})&\stackrel{(a)}{\le} 
\sum_{i\le\trans}\wred(\cPPsns{nf_i})+\sum_{i>\trans}\wred(\cPPsns{nf_i})\\
&\stackrel{(b)}{\le} \sum_{i\le\trans} \log\left(2+\sqrt{\frac{2
nf_i}{\pi}}\right)+\Big(\sum_{i>\trans}^{\infty} {nf_i}\Big)\log e,
\end{align*}
where $(a)$ follows from Theorem~\ref{thm:env_class} and
$(b)$ from Lemma~\ref{lem:one_poisson}.

For the first term, note that for $\Lambda\ge1$, 
$2+\sqrt{2\Lambda/\pi}<3\sqrt{\Lambda}$.
Therefore,
\begin{align*}
\sum_{i=1}^{b}\log \frac{b}{i} = \log \frac{b^b}{b!} \le \log e^{b}=
b\log e,
\end{align*}
 where the inequality follows from Stirling's approximation (Equation~\eqref{eqn:stirling}). 
Using $b=(\const n)^{\frac1\alpha}$, 
\begin{align*}
\sum_{i=1}^{\trans} \log\left(2+\sqrt{\frac{2
\upper_i}{\pi}}\right)
&< \sum_{i=1}^{\trans}\log \left(3\sqrt{\frac{\const n}{i^{\alpha}}}\right)\\
&= \trans\log (3) + \frac\alpha2 
 \sum_{i=1}^{b}\log \Big(\frac {b}{i}\Big)\\
&< (\const n)^{1/\alpha}\Big(\log (3) + \frac{\alpha\log e}2\Big).
\end{align*}

For the second term,  
\begin{align*}
\sum_{i=\trans+1}^{\infty} nf_i
= \const n\sum_{i=\trans+1}^{\infty}\frac1 { i^{\alpha}}
<\const n\int_{\trans}^{\infty}\frac{dx}{x^{\alpha}} 
=\frac{\const^{1/\alpha}}{\alpha-1}n^{1/\alpha}.
\end{align*}

Using $\wred(\Pwrcalpn)\le \wred(\Pwrcalp^{\Psnn})+1$
from Theorem~\ref{thm:poisson_fixed} and adding the terms
proves the upper bound.

\subsubsection{Lower Bound}

The sum 
\[
\sum_{j=\ell+1}^{\infty}f_i = \sum_{j=\ell+1}^{\infty}\frac c{i^{\alpha}}\le
\int_{\ell}^{\infty}\frac{c}{x^{\alpha}}dx = \frac{c\ell^{-\alpha+1}}{\alpha-1}
\]
is less than 1 for any $\ell > \ell_0\ed\left(\frac
  {\const}{\alpha-1}\right)^{\frac1{\alpha-1}}$. 

Hence by Theorem~\ref{thm:env_class}, 
\begin{align*}
\wred(\Pwrcalp^{\Psnn}) \ge \sum_{i> \ell_0}\wred(\cPPsns{nf_i}) 
=
\sum_{\ell< i\le \trans}\wred(\cPPsns{nf_i}) + \sum_{i>\trans}\wred(\cPPsns{nf_i}),
\end{align*}
where recall that $\trans = (\const n)^{1/\alpha}$.

For the first term, using Lemma~\ref{lem:one_poisson}, 
\[
\sum_{\ell_0< i\le \trans}\wred(\cPPsns{nf_i})
\ge \sum_{\ell_0< i\le \trans}\log \sqrt{\frac{2nf_i+2}{\pi}}
\ge \frac 12\sum_{\ell_0< i\le \trans}\log \frac{2n\const}{\pi i^{\alpha}}
> \sum _{1\le i\le \trans} \log \frac{n\const}{i^{\alpha}}-\ell_0\log
(nc)-\frac \trans 2 \log \frac \pi 2.
\]
Again by Stirling approximation, and using $B!<2\pi B (B/e)^B$, 
\[
\sum _{1\le i\le \trans} \log \frac{n\const}{i^{\alpha}}
=\log \frac {(cn)^b}{(b!)^\alpha}
\ge \frac \alpha 2 (cn)^{1/\alpha}\log e -\alpha\log (2\pi b)
\]

For the second term, using $2-e^{-\lambda}>1+\lambda/2$ for
$\lambda<1$, 
\begin{align*}
\sum_{i>\trans}\wred(\cPPsns{nf_i}) 
=\sum_{i>\trans}\log (2-e^{-nf_i})
>\sum_{i>\trans}\frac{nf_i}2\log e
>\frac{\const n}2\sum_{i>\trans}\frac1{i^{\alpha}}
> (\const n)^{1/\alpha}\frac1{2(\alpha-1)}-O(1).
\end{align*}

Summing these two lower bounds $\wred(\Pwrcalp^{\Psnn})$. 
Invoking the lower bound of Theorem~\ref{thm:poisson_fixed}
with the upper bound from the previous part proves the lower 
bound. We hide the lower order and logarithmic terms in the 
$o(1)$ factor. 
\end{proof}

\begin{Remark}
From a simple calculation, it follows that the two bounds are 
at most a factor of 4 apart. 
By obtaining tighter bounds on $\wred(\cPPsnupr)$, it should be 
possible to obtain upper and lower bounds within a multiplicative
$(1+\delta)$ factor for arbitrarily small $\delta$. 
\end{Remark}

\section{Exponential Envelopes}
We provide a simple proof of Equation~\eqref{eqn:bontemp_exponential}
with stronger second order terms.
More precisely, we prove that
\begin{Theorem}
\[
 \wred(\Expcalpn) = \frac{\log^2 n}{4\alpha}+O(\log \const\log n).
\]
\end{Theorem}
\begin{proof}
For the exponential class, $f_i = \const e^{-\alpha i}$. Therefore,
\[
i\le \frac{\ln (\const n)}{\alpha}\ \ \Leftrightarrow \ \ nf_i\ge 1.
\] 
Similar to the argument for power-law, let $b\ed \frac{\ln (\const
  n)}{\alpha}$, and by Theorem~\ref{thm:env_class} and
Lemma~\ref{lem:one_poisson}, 
\begin{align*}
\wred(\Expcalp^{\Psnn})
\le 
\sum_{i\le\trans}\wred(\cPPsns{nf_i})+\sum_{i>\trans}\wred(\cPPsns{nf_i})
\le  \sum_{1\le i\le b} \log\Big(2+\sqrt{\frac{2
nf_i}{\pi}}\Big)+(\sum_{i>b} nf_i)\log e
\end{align*}
Using $e^{b\alpha} = \const n$, the second term can be bounded using
\[
\sum_{i>b}nf_i= \sum_{i>b}
\const ne^{-\alpha i}< \const ne^{-\alpha b}\frac1{1-e^{-\alpha}}
= \frac1{1-e^{-\alpha}}.
\]
Following the same steps as power-law and using
$e^{b\alpha}=n$, 
\begin{align*}
\sum_{i=1}^{b}\log \left(2+\sqrt{{\frac{2nf_i}\pi}}\right)
&\le \sum_{i=1}^{b}\log \Big(3\sqrt{nf_i}\Big)\\
&\leq  b\log {3} + \frac 12 \sum_{i=1}^{b}\log [n\cdot\const e^{-\alpha i}]\\
&\leq  b\log 3 +\frac 12\log [(\const n)^b\cdot(\const n)^{-(b+1)/2}]\\
&= b\log 3 +\frac {(b-1)}4 \log (\const n)\\ 
&< \frac b4\log (81\const)+ \frac b4 \log n.
\end{align*}
Substituting $b=\ln (\const n)/\alpha$,  
\[
\wred(\Expcalpn)\le \frac{\log^2 n}{4\alpha\log e} + \frac{\log cn\log
  (81c)}{4\alpha\log e}+\frac{\log n \log c}{4\alpha\log e}+\frac{\log e}{1-e^{-\alpha}}+1.
\]
We now prove the lower bound by a similar argument. Clearly, for $\ell\geq\ell_0\ed
\frac1\alpha \log (\frac \const{1-e^{-\alpha}})$, 
\[
\sum_{i=\ell}^{\infty}f_i\leq1. 
\]
Recall that $b=\ln (\const n)/\alpha$, then by Lemma~\ref{lem:one_poisson},
\begin{align*}
\wred(\Expcalppsnn)&\ge \sum_{i=l_0}^{b}\log \left(\sqrt{\frac{2nf_i+2}{\pi}}\right)\\
&\ge \frac 12 \Big[ \sum_{i=1}^{b}\log \frac{2nf_i}{\pi}
- \sum_{i=1}^{\ell_0}\log \frac{2nf_i}{\pi}\Big]\\
&\ge \frac 12\Big[b\log \frac 2\pi +\log [(cn)^b\cdot(cn)^{-(b+1)/2}]-
\sum_{i=1}^{\ell_0}\log \frac{2nf_i}{\pi}\Big]\\
&\ge \frac 12\Big[b\log \frac 2\pi +\frac{b-1}2\log (cn)-
\sum_{i=1}^{\ell_0}\log \frac{2nf_i}{\pi}\Big]\\
&\ge \frac {b-1}4\log \frac{4cn}{\pi^2}-\frac{\ell_0}4\log
\frac{4cn}{\pi^2}\\
 &\ge \frac{\log^2 n}{4\alpha\log e} -O(\log \const\log n).
\end{align*}
To obtain a bound on fixed length sampling, we apply
Theorem~\ref{thm:poisson_fixed} with the upper bound proved
earlier. 
\end{proof}

\section{A note on expected redundancy}
Since the worst case redundancy is always larger than the average
redundancy, a class with infinite average redundancy has infinite
worst case redundancy. 
However, there exist classes of distributions with a finite average
redundancy and infinite worst case redundancy (See Example 1
in~\cite{AcharyaDJOS13}).

However, in Lemma~\ref{lem:worst_avg} we show that an envelope class
with infinite worst case redundancy also has infinite average case
redundancy. On a related note,~\cite{BGG09} showed that for any
distribution class with finite worst case redundancy, the worst case
redundancy grows sub-linearly with $n$, namely if $\wred(\cP)<\infty$, 
then $\wred(\cP)=o(n)$. Recently,~\cite{HosseiniS14} showed that there
exists a class $\cP$ such that $\ered(\cP)<\infty$, and yet,
$\ered(\cP^n)=\Omega(n)$, \ie the redundancy grows linearly with the
block length. 

\begin{Lemma}
\label{lem:worst_avg}
For any envelope function $f$, 
\[
\wred(\Envcls) < \infty\Leftrightarrow \ered(\Envcls)<\infty.
\]
\end{Lemma}
Our proof of this lemma uses the following result.
\begin{Lemma}
If $\dclass$ contains $M$ distributions over mutually
disjoint supports, then $\ered(\dclass)\ge\ln M$. 
\end{Lemma}
\begin{proof}
Let $\distP_1, \ldots, \distP_M$ be $M$ distributions over disjoint
sets $\cA_1, \ldots,\cA_M$. For any distribution $Q$, there
is a $j$, such that $\distQ(\cA_j)\le 1/M$. For that distribution, 
\[
D(\distP_j||\distQ) =\sum_{x\in\cA_j}P_j(x)\ln \frac{P(x)}{Q(x)} \ge -\ln
\left(\sum_{x\in\cA_j}\distQ(x)\right)\ge \ln M,
\]
where the inequality is from the convexity of logarithms.
Since this holds for any distribution $\distQ$, it holds for the 
infimum and plugging in the definition of $\ered$ proves
the result.
\end{proof}

\begin{proof}[Proof of Lemma~\ref{lem:worst_avg}]
The forward direction is trivial since $\ered\le\wred$. For the 
other direction, we show that if $\wred(\dclass)=\infty$, there
exist an infinite number of distributions $\distP_1, \distP_2,\ldots$
that all have disjoint supports, and applying the previous lemma
to this proves the result. Now, $\wred(\dclass)=\infty$ means
$\sum f_i=\infty$. Using this we construct an infinite sequence
of distributions as follows. After constructing distribution $\distP_i$
consider the first integer not in its support, and construct a 
distribution over the smallest possible integers starting from 
that location. This process can be repeated infinitely many times
giving an infinite number of distributions with disjoint supports. 
\end{proof}

\footnotesize{\bibliographystyle{plain}
\bibliography{isit}}
\appendix
\end{document}